# Spectroscopic study of five Mira stars

## David Boyd

Mira stars are popular targets for amateur astronomers because they are generally bright and their pulsations produce large changes in magnitude. This also means they are eminently suitable to study using spectroscopy. We report here on spectroscopic observations of five Mira stars: RY Cep, SU Cam, T Cep, V667 Cas, and OR Cep. We describe the life cycle of a Mira star and explain how the carbon-oxygen balance in its atmosphere determines its spectral classification. Atmospheric pulsations in these stars produce shocks which generate Balmer emission. We present graphically their different spectral behaviours and development of emission lines.

## Introduction to Mira stars

Mira stars are pulsating red-giant stars with masses between a half and around eight solar masses, radii over 100 times that of the Sun, spectral type late K or M, luminosity class III, pulsation periods of several hundred days, and visual amplitudes of more than 2.5 magnitudes. They are evolved stars approaching the end of their active thermonuclear lives and are named after the prototype of this type of star, omicron Ceti (Mira).

A typical Mira star begins life on the main sequence of the Hertzsprung–Russell (HR) diagram, burning hydrogen into helium in its core and generating sufficient outward radiation pressure to counteract the continual inward pull of gravity (Figure 1).

Eventually, after billions of years, most of the hydrogen in the core has converted to inert helium and the core has cooled. As it is no longer producing enough radiation to resist gravity, it begins to collapse. This generates enough heat to ignite a shell of remaining hydrogen above the core. Electron degeneracy in the core eventually counteracts gravity and stops further contraction.

Heat released by hydrogen burning causes the outer atmosphere of the star to expand and cool. The star is leaving the main sequence, becoming a subgiant and starting to climb towards the upper-right of the HR diagram along the red giant branch (RGB). As more helium is added to the core, its temperature rises and the outer envelope of the star expands further. Towards the top of the RGB, helium in the core starts burning to form first carbon and then oxygen by fusion of helium nuclei in the triple-alpha process. In lower-mass stars with a relatively small core, this process happens dramatically with a helium flash. As helium burns in the core and hydrogen continues to burn in a shell around the core, the star travels back along the horizontal branch (HB) with approximately constant luminosity.

Eventually, helium in the core is exhausted, leaving the star with a small degenerate carbon-oxygen core surrounded by a helium-burning shell and, above that, a hydrogen-burning shell. These shells tend to burn alternately, with a decrease in one leading to an increase in the other. The star's outer atmosphere has now expanded to a radius of 1 au or more and the star is in the asymptotic giant branch (AGB). It continues to experience helium shell flashes which can generate thermal pulses lasting hundreds or thousands of years. Convection dredges up carbon and other products of nuclear burning, including heavier s-process elements formed by slow neutron capture, into the star's outer layers where they may manifest themselves in the star's spectrum. The proportion of these different elements in the atmosphere of the giant star determines its chemical and physical nature.

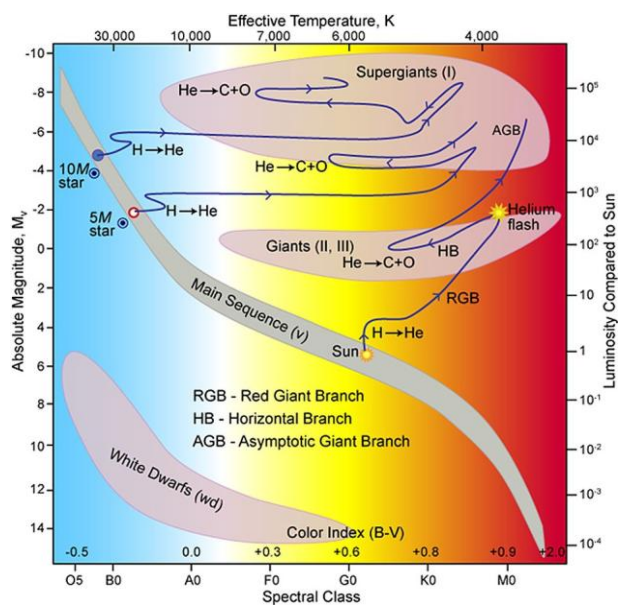

**Figure 1.** Hertzsprung–Russell diagram showing the path of an evolving Mira star. *(Robert Hollow, Commonwealth Science and Industrial Research Organisation (CSIRO), Australia, adapted by Carin Cain)*

Meanwhile, competition between radiation pressure and the pull of gravity in the star's atmosphere generates regular pulsations, typically lasting a year, which produce the characteristic light curve of a Mira variable.

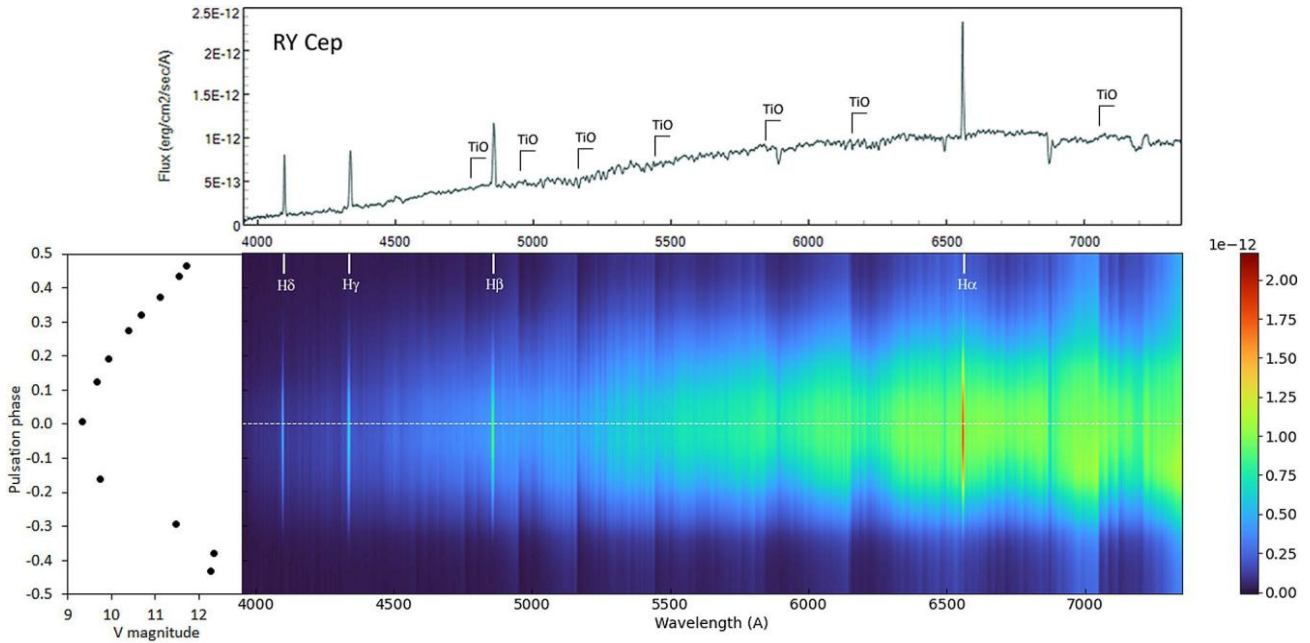

**Figure 2.** Spectrum of RY Cep at the peak of the pulsation cycle, with a plot of its changing V magnitude and a 2-D image showing the progressive change in absolute flux (colour-coded in erg/cm²/s/Å) with pulsation phase during the cycle. The dotted white line marks the phase of the pulsation peak. The main molecular absorption bands are labelled in the spectrum and the position of the hydrogen Balmer lines are marked at the top of the 2-D image.

This final phase of the star's life is relatively short. An increasingly strong stellar wind progressively expels the convective envelope and the atmosphere of the star. In a relatively small number of cases, a planetary nebula may be formed. The remaining degenerate carbon-oxygen core becomes a very hot but slowly cooling white dwarf.

This is an oversimplified description of complex and not-yet-fully-understood behaviour. A readable review of our knowledge of Mira stars is given in Willson & Marengo (2012).[1]

## The carbon/oxygen ratio

Miras can be divided into M-type, S-type, and carbon stars according to the relative abundance of oxygen and carbon in their atmospheres. This is often characterised in terms of a carbon/oxygen or C/O ratio. The atmosphere of a giant star is sufficiently cool that molecules form. The relative abundance of different elements in the atmosphere determines the chemical composition of these molecules and the spectral classification of the star.

– If there is more oxygen than carbon in the outer layers of the star (C/O < 1), it is described as an oxygen-rich M giant. Most of the carbon is tied up in carbon monoxide (CO) molecules. Molecular oxides of titanium and other metals form in the star's atmosphere and absorb light in specific wavelength bands, creating a characteristic saw-tooth spectral profile.

– If oxygen and carbon are present in approximately equal amounts (C/O ~ 1), the star is referred to as an S-type or S star. These stars have absorption lines and bands in their spectra due to the presence of s-process elements such a strontium, yttrium, barium and zirconium, with bands of zirconium oxide being the signature of S stars.

– If carbon predominates (C/O > 1), it is called a carbon star and the atmosphere contains compounds of carbon such as CO, which will have absorbed all the available free oxygen, diatomic carbon ($C_2$), cyanogen (CN) and methylidyne (CH). Carbon in the star's atmosphere tends to hide the absorption features caused by other molecules.

As a Mira star slowly climbs the AGB, repeated dredge-ups can increase the C/O ratio, changing the star from an oxygen-rich M-type to an S-type and possibly eventually a carbon star.

## Observations

The Mira stars we report on here are RY Cep, SU Cam, T Cep, V667 Cas and OR Cep. By choosing stars at high declination, it has been possible to follow them through their year-long pulsation cycles using both spectroscopy and photometry, the latter obtained concurrently with the spectra. Table 1 lists the periods of these pulsation cycles and the epochs corresponding to phase zero (maximum light) of the cycles. Low-resolution ($R \sim 1000$) spectra were obtained with a LISA spectrograph on a 0.28-m C11 Schmidt–Cassegrain telescope (SCT) and photometry obtained with a 0.35-m SCT equipped with BVRcIc filters. Spectra are calibrated in absolute flux using the concurrently obtained V magnitudes, as described in Boyd (2020).[2] Further information on these observations and how they were analysed can be found in Boyd (2021),[3] Boyd (2023),[4] and Boyd (2024).[5]

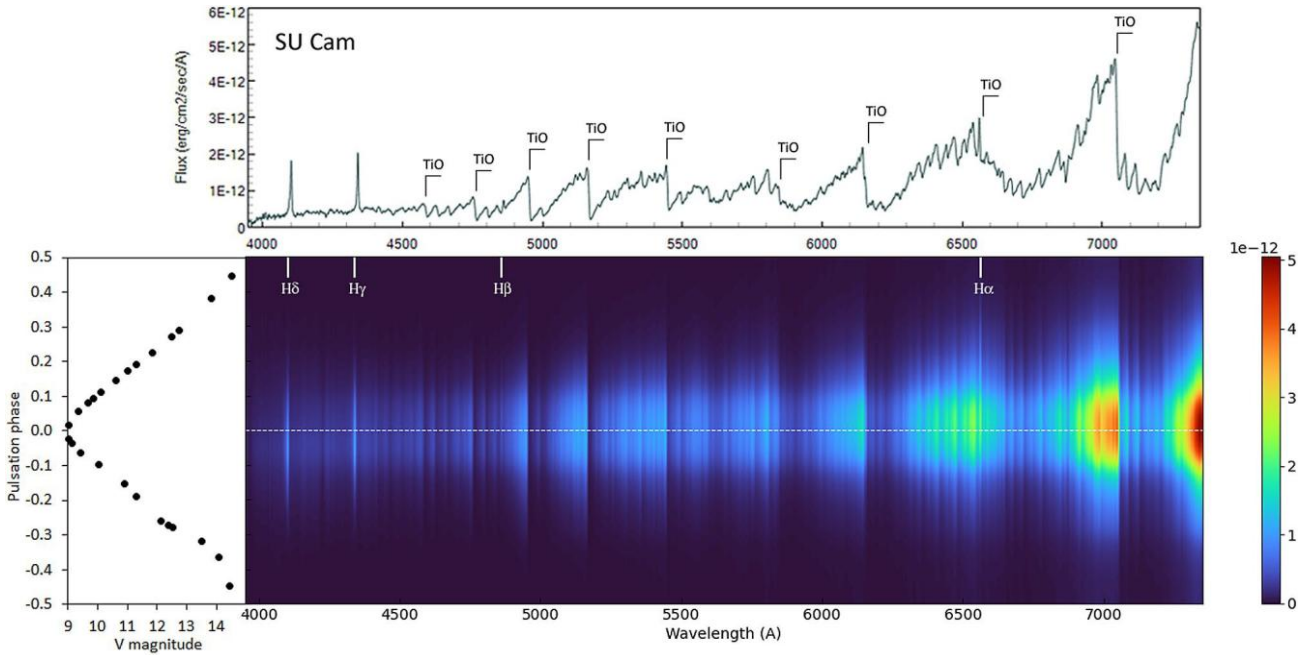

**Figure 3.** Spectrum of SU Cam. Details as for Figure 2.

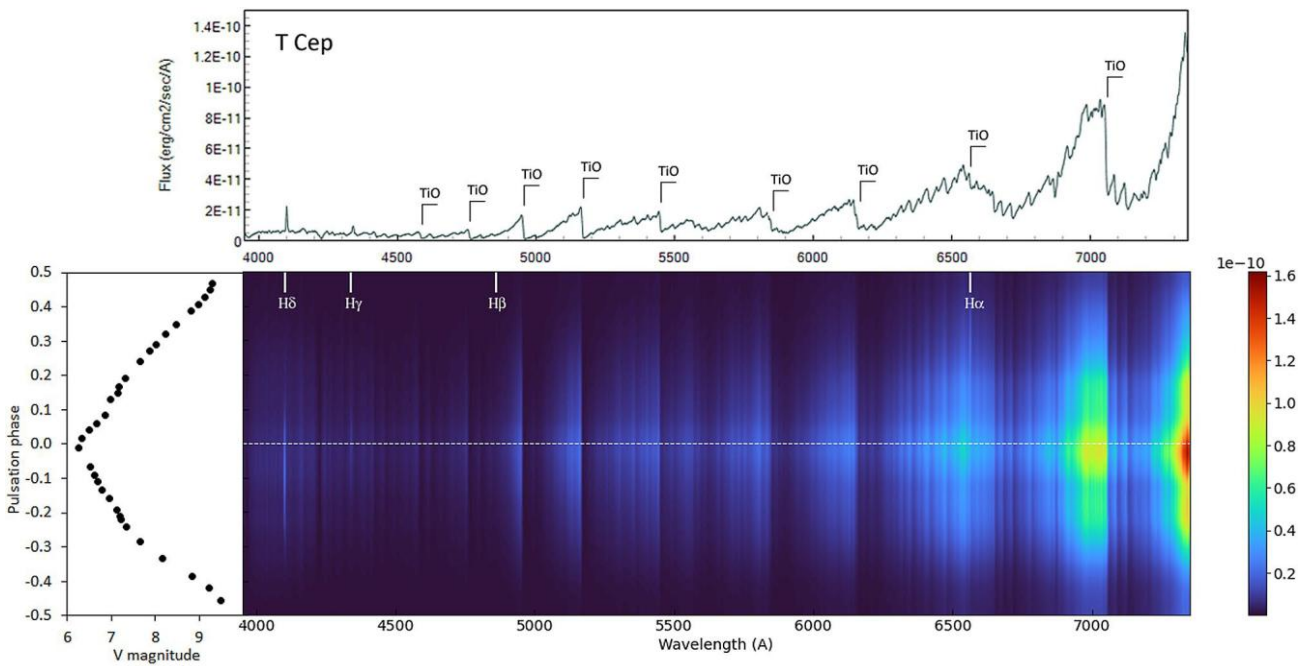

**Figure 4.** Spectrum of T Cep. Details as for Figure 2.

Figures 2–6 show the spectrum of each star at the peak of the pulsation cycle, with a plot of its changing V magnitude and a 2-D image showing the progressive change in absolute flux (colour-coded in erg/cm$^2$/s/Å) as it passed through a complete pulsation cycle. The dotted white lines mark the phase of the pulsation peak. The main molecular absorption bands are labelled and the wavelengths of the hydrogen Balmer lines are marked at the top of the 2-D plot. The colour-coding of the 2-D plots shows the progressive change in the profile of the spectra during a pulsation cycle. Data analysis and graphical output used custom Python software written by the author.

**Table 1. Measured pulsation cycle period & epoch of maximum light (phase zero) of that cycle**

| Star | Pulsation period (d) | Epoch of max. light (JD) |
|---|---|---|
| RY Cep | 153 | 2458798 |
| SU Cam | 292 | 2458268 |
| T Cep | 387 | 2459747 |
| V667 Cas | 340 | 2459758 |
| OR Cep | 347 | 2459685 |

## Spectral type

Assigning a specific spectral type to a giant star is not easy because of the wide variety of behaviour encountered in those stars. It is also difficult with low-resolution spectra, in which many of the distinguishing features used in spectral classification are not well resolved. In our case, we adopted the approach of finding the statistically best morphological fit of the spectra of our stars with those of stars already classified on the revised MK system and its extensions (Gray & Corbally, 2009).[6]

For those stars described in *SIMBAD* (Wenger, 2000) as M-type,[7] namely RY Cep, SU Cam and T Cep, we compared our spectra with a range of spectral types from the Perkins (Keenan & McNeil, 1989) and Fluks (Fluks *et al.*, 1994) catalogues,[8,9] and interpolated between their closest-fitting spectral types to assign values to the nearest decimal point. For the S-type star V667 Cas, we assigned a spectral type based on the spectral classification criteria on the revised MK system in Keenan & Boeshaar (1980).[10] This depends on comparing the relative strengths of TiO and ZrO molecular bands and the Na I D absorption lines. For the carbon star OR Cep, we assigned a spectral type based on the closest fit with spectra in the *Spectral Atlas of Carbon Stars* in Barnbaum *et al.* (1996),[11] which are classified according to the revised MK system for carbon stars in Keenan (1993).[12]

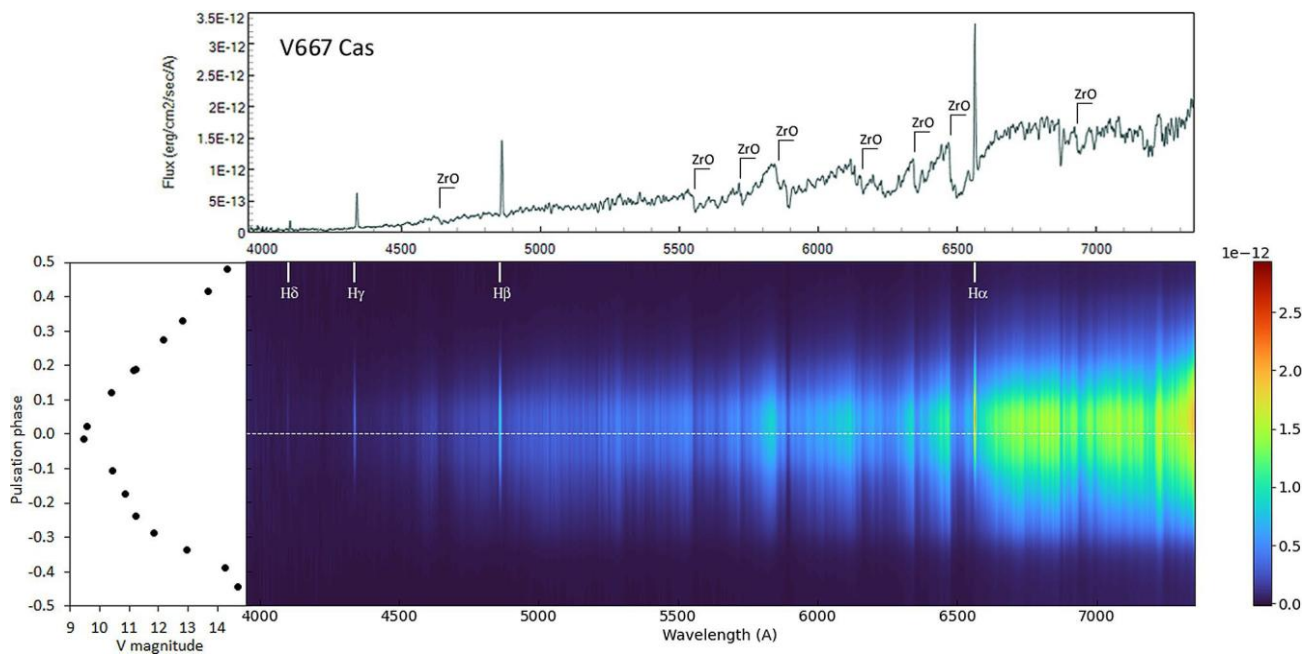

**Figure 5.** Spectrum of V667 Cas. Details as for Figure 2.

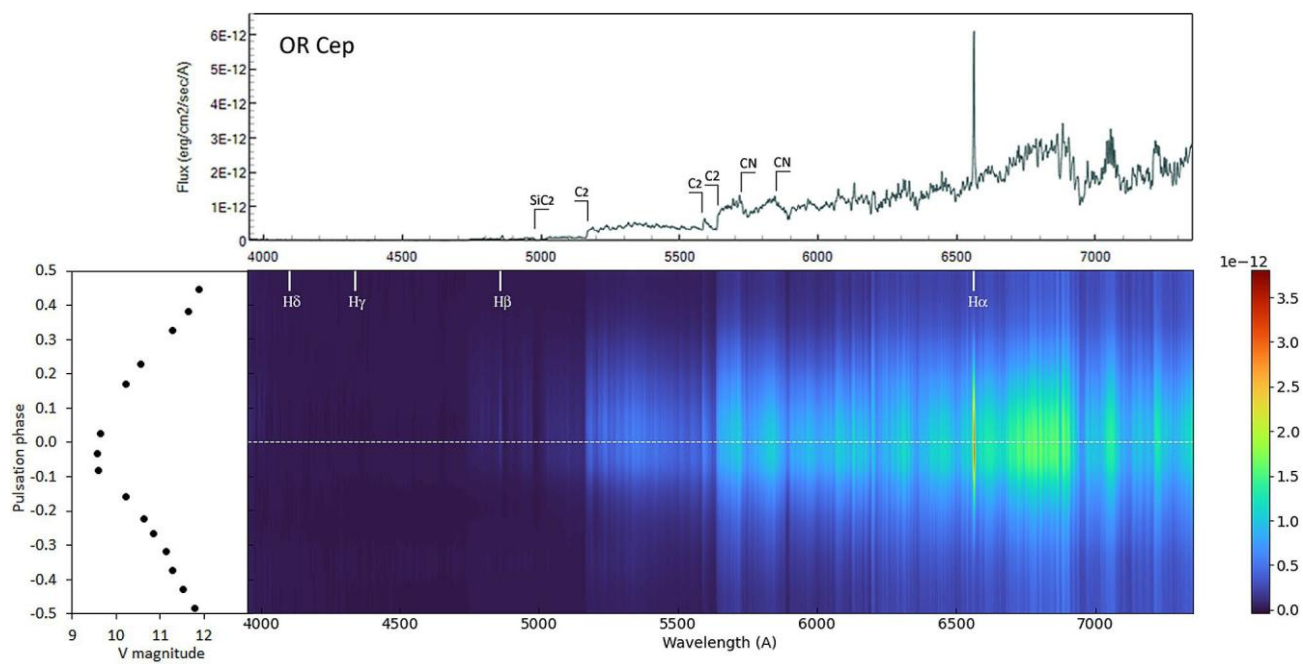

**Figure 6.** Spectrum of OR Cep. Details as for Figure 2.

In Table 2, we list the spectral type assigned to each star at the peak of its pulsation cycle and at minimum. As a Mira fades and becomes cooler during each cycle, molecules reform, causing the absorption bands to deepen and the apparent spectral type to become later. At maximum, RY Cep is a K-type star and shows almost no absorption bands, but it becomes M-type at minimum with weak TiO bands. SU Cam and T Cep are M-type stars throughout their cycles and show prominent titanium oxide (TiO) bands. V667 Cas is an S-type star and the absorption bands are of zirconium oxide (ZrO). The flux of V667 Cas was so low at minimum that an assessment of spectral type was not possible. OR Cep is a carbon star and the absorption is due to various carbon molecules.

**Table 2. Spectral types assigned to each star at maximum and minimum of the pulsation cycle**

| Star | Spectral type at max. | Spectral type at min. |
| --- | --- | --- |
| RY Cep | K4.2 | M6.5 |
| SU Cam | M5.5 | M8.6 |
| T Cep | M7.5 | M9.4 |
| V667 Cas | SX/6 | – |
| OR Cep | C-N5+ | C-J4.5 |

For V667 Cas, the flux at minimum was so low that determining a spectral type was not possible.

## Pulsation, shocks & Balmer emission

The mechanism driving pulsation in Mira stars is still a subject of debate. The kappa mechanism explains pulsation in terms of changing opacity. One possible cause is a cycle of ionisation and recombination of hydrogen in a layer inside the photosphere. Ionisation increases opacity and pressure of radiation causes the layer to expand outwards. As it cools, the hydrogen recombines, opacity drops and gravity pulls the layer inwards. Another possibility is that changing opacity could be caused by a cycle of dust formation and growth in the atmosphere. These repeating pulsation cycles in the atmosphere of the star drive mass loss.

The transient appearance of hydrogen Balmer emission lines in the spectra of Mira stars was first detected in early objective-prism observations at the Harvard College Observatory (Maury & Pickering, 1897).[13] They were seen to appear as the star approached maximum light and disappear as it faded, although their cause was not understood. It was Gorbatskii (1961) who first explained theoretically that shock waves generated deep in the star's atmosphere could be the cause of the emission lines seen in Mira and other stars.[14] These shocks form during the pulsation cycle, when gas falling inward from the previous cycle meets gas pushing outward in the next. This ionises hydrogen in the region behind the shock, which then recombines following the passage of the shock and generates emission lines. These lines strengthen as the pulsation cycle grows, eventually rising above the spectral continuum and reaching maximum strength around the peak of the cycle.

The absolute flux in each Balmer emission line in each spectrum was measured by integrating the profile of the line above a local continuum interpolated under the line. Figures 7–11 show how the integrated flux in erg/cm$^2$/s in four hydrogen Balmer lines varied during the pulsation cycle of each star. In T Cep, emission is double-peaked, with a different balance between peaks for each line; this behaviour is not seen in the other stars. Mira stars with later spectral types have cooler atmospheres which contain more molecules in their upper layers. These absorb energy from emission lines generated lower down, thus reducing their strength. This alters the normal Balmer line decrement (H$\alpha$ > H$\beta$ > H$\gamma$ > H$\delta$) by progressively suppressing longer-wavelength lines in later spectral types. This is evident comparing Balmer lines in the three M-type stars. As the C/O ratio increases, this pattern reverses and it is the shorter-wavelength Balmer lines which are progressively suppressed.

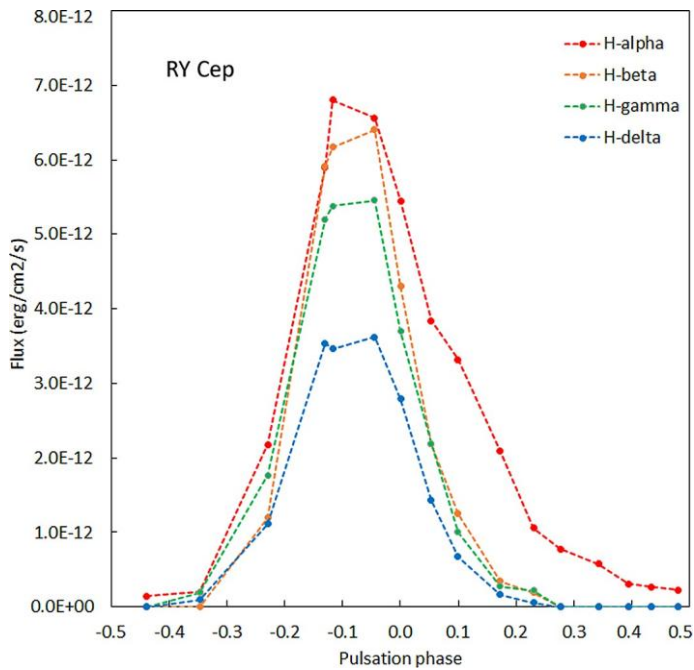

**Figure 7.** Variation of the integrated flux of the Balmer lines in RY Cep during a pulsation cycle.

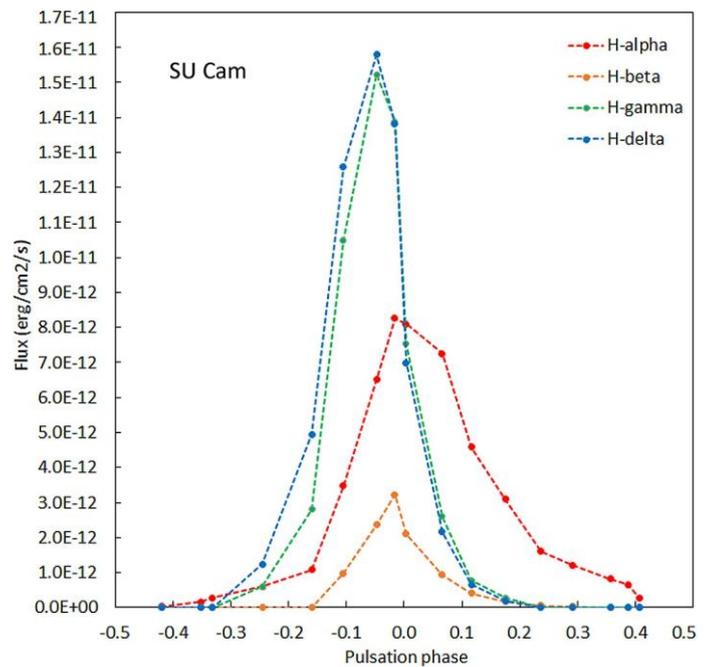

**Figure 8.** Variation of the integrated flux of the Balmer lines in SU Cam during a pulsation cycle.

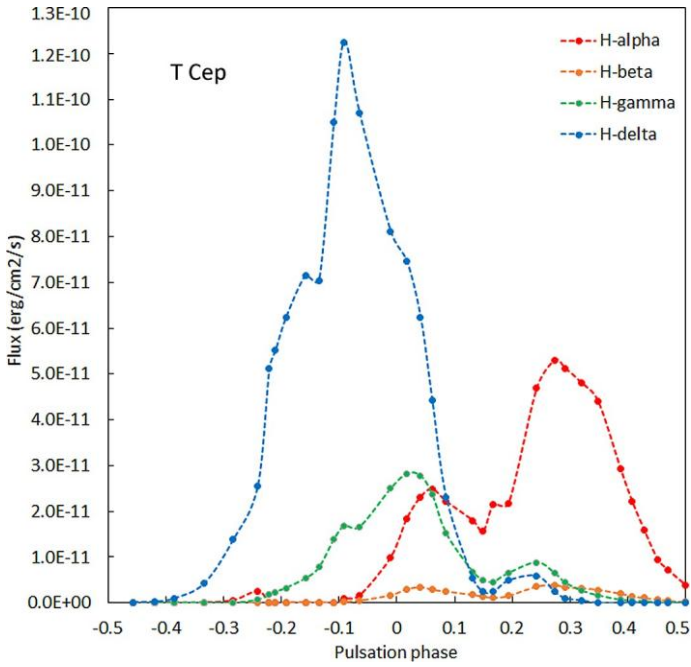

Figure 9. Variation of the integrated flux of the Balmer lines in T Cep during a pulsation cycle.

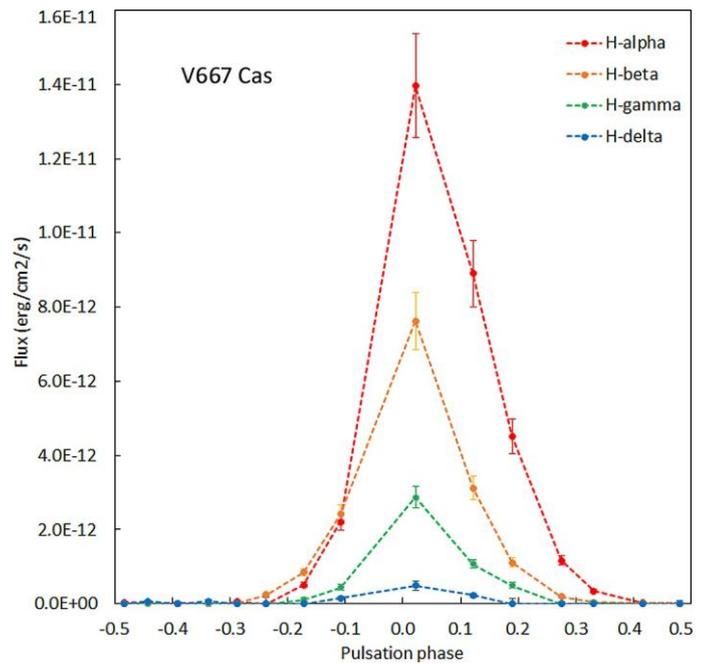

Figure 10. Variation of the integrated flux of the Balmer lines in V667 Cas during a pulsation cycle.

## Summary

Using a combination of low-resolution spectroscopy and V-band photometry we made quantitative measurements of the spectra of five Mira stars and presented graphically how their spectral profiles, V magnitudes and Balmer emission lines changed during a pulsation cycle.

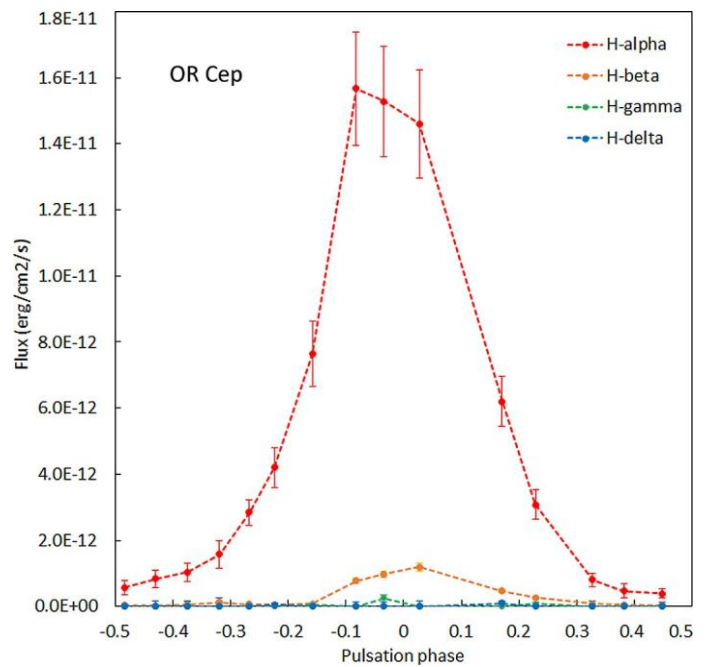

Figure 11. Variation of the integrated flux of the Balmer lines in OR Cep during a pulsation cycle.

## Acknowledgements

We acknowledge with thanks the advice and encouragement of Lee Anne Willson. This research made use of the SIMBAD database, CDS, Strasbourg Astronomical Observatory, France (Wenger, 2000),[7] and the AAVSO Photometric All-Sky Survey (Henden 2021).[15] We are also grateful to the developers of the *Astropy* package and other contributors to this valuable community resource (Astropy Collaboration *et al.*, 2018).[16]